\documentclass[aps,prl,twocolumn,showpacs]{revtex4}
\usepackage{amssymb}
\usepackage{amsmath, amsthm}
\newcommand{\be}{\begin{equation}}
\newcommand{\ee}{\end{equation}}

\begin{document}


\title{The correspondence between a scalar field and an 
effective perfect fluid} 


\author{Valerio Faraoni}
\email[]{vfaraoni@ubishops.ca}
\affiliation{Physics Department and STAR Research Cluster, 
Bishop's University, 2600 
College Street, 
Sherbrooke, Qu\'ebec, Canada J1M~1Z7
}

\begin{abstract} 
It is widely acknowledged that, for formal purposes, a 
minimally coupled scalar field is equivalent to an 
effective perfect fluid 
with equation of state determined  by the scalar potential.
This correspondence is not complete because the Lagrangian 
densities  ${\cal L}_1=P$ and ${\cal L}_2=-\rho$, which 
are equivalent  for a perfect fluid, are not equivalent for 
a minimally coupled scalar field. The exchange ${\cal L}_1 
\longleftrightarrow {\cal L}_2$ amounts to exchanging a 
``canonical'' scalar field with a phantom scalar field.
\end{abstract}

\pacs{04.40.-b, 04.20.Fy, 04.20.-q}

\maketitle

\section{Introduction}
\setcounter{equation}{0}

Scalar fields have been present in the literature on 
gravity and cosmology for many decades  and are 
particularly important in 
the contexts of inflation in the early universe and of 
quintessence in the late universe.  It is widely 
recognized that a minimally coupled scalar 
field in General Relativity can be represented as a perfect 
fluid (see, {\em e.g.}, Ref.~\cite{Madsen} for a detailed 
discussion). From the conceptual point of view, it is 
clear that a scalar field and a perfect fluid are very 
different physical systems; a fluid is obtained by 
averaging 
microscopic quantities associated with its constituent 
particles, and the fluid laws can be obtained via a  
kinetic theory using microscopic models of the fluid 
particles and of their interactions.  A scalar field, 
instead, 
is more fundamental and does not derive from an average; it 
is not made of point-like particles and does not need an 
average to be defined---it is already a continuum. As a 
consequence, the scalar field stress-energy tensor (given 
by eq.~(\ref{tmunu}) below)  
involves first order derivatives of 
the scalar field, while the perfect fluid stress-energy 
tensor $T_{ab}=\left(P+\rho \right) u_a u_b +Pg_{ab}$ does 
not involve derivatives explicitly but can be 
described using only the energy density $\rho$, pressure 
$P$, and four-velocity field $u^a$. Hence, the fluid 
description of a scalar field is not an identification, 
but only a convenient correspondence for formal 
purposes. However, while regarding the scalar field/perfect 
fluid duality as a  formal one, one still 
needs to be careful because this correspondence is not 
perfect even for purely formal purposes. In this short note  
we discuss a property of perfect fluids which is not 
satisfied by 
the effective fluid associated with a scalar field, namely 
the possibility of describing a perfect fluid with the two 
equivalent Lagrangian densities ${\cal L}_1=P$ and ${\cal 
L}_2=-\rho$. These two Lagrangian densities, which are 
equivalent for a perfect fluid \cite{SchutzBrown, 
HawkingEllis}, are not equivalent for a 
scalar field ``fluid'', as shown in the following section. 
This property leads to a caveat against regarding the 
fluid/scalar field duality as  a perfect one, which is 
relevant in the light of the recent interest in this 
correspondence motivated by the possibility of computing 
inflationary perturbations on one side of the duality by 
using the other side (\cite{Boubekeretal, ArrojaSasaki}, 
see also \cite{UnnikrishnanSriramkumar}).

The non-equivalence between these two Lagrangian densities 
for a perfect fluid coupling explicitly to the Ricci 
curvature of spacetime (which has been suggested as 
a possible alternative to dark 
matter  in galaxies \cite{BBHL}) has been recently 
discussed in \cite{BBHLinequivalence}, and 
the equivalence between a k-essence scalar field and a 
barotropic perfect fluid was discussed in 
\cite{ArrojaSasaki}.

We adopt the notations of Ref.~\cite{Wald}; in particular, 
the signature of the spacetime metric is $-+++$.

\section{Mimicking a perfect fluid with a minimally coupled 
scalar field}

Let us mimick a perfect fluid using a minimally coupled 
scalar 
field $\phi$ in a  curved spacetime self-interacting 
through a 
potential $V(\phi) $. We assume standard 
General Relativity and the issue is to determine whether  
both Lagrangian densities ${\cal L}_1=P$ and 
${\cal L}_2=  -\rho$ correctly describe the scalar field 
effective fluid.  It is well known that these Lagrangian 
densities are equivalent for a perfect fluid 
\cite{SchutzBrown}.

A minimally coupled scalar field $\phi$ obeys the 
Klein-Gordon equation
\be\label{KleinGordon}
\Box \phi -\frac{dV}{d\phi}=0
\ee
which is obtained (when $\nabla^c\phi $ does not vanish 
identically) from the covariant conservation equation $\nabla^b 
T_{ab}=0$ for the scalar field energy-momentum tensor
\be \label{tmunu}
T_{ab} [\phi]=\nabla_a\phi \nabla_b\phi
-\frac{1}{2}\, g_{ab} \nabla^c \phi \nabla_c\phi -V g_{ab} \,.
\ee
The tensor $T_{ab} [\phi]$   assumes the form of  an 
effective  perfect fluid 
stress-energy tensor $T_{ab}=\left( P+\rho \right) u_a u_b +P 
g_{ab} $ if $\nabla^c \phi$ is a timelike vector field 
\cite{Madsen}. The  fluid four-velocity is
\be 
 u_a  =  \frac{ \nabla_a \phi}{ \sqrt{ \left| \nabla^c \phi 
\nabla_c \phi \right| } } \label{4velocity}
\ee
assuming $ \nabla^c \phi \nabla_c \phi \neq 0$, with $ u_a 
u^a =  \mbox{sign} \left( \nabla^c \phi 
\nabla_c \phi \right)$. The  energy density and pressure 
relative to a comoving observer with four-velocity $u^a$  
are given by $ \rho = T_{ab}[\phi ] u^au^b$ and 
$ P= T_{ab} [\phi] h^{ab} /3$, respectively, 
where  $ h_{ab }  \equiv g_{ab }  + u_a u_b  $ 
is the Riemannian 3-metric in the 3-space orthogonal to 
$u^c $ (this $3+1$ decomposition makes sense when 
$\nabla_c \phi$ is timelike). One easily obtains
\begin{eqnarray}
\rho &=& \left( \frac{1}{2} \, \nabla^c\phi \nabla_c \phi 
-V  \right) \mbox{sign} \left( \nabla^c\phi \nabla_c\phi 
\right) \,, \label{questa1} \\
&& \nonumber\\
P &=& \frac{1}{3} \left\{ \left[ -1+\frac{1}{2} 
\, \mbox{sign} 
\left( \nabla^c\phi \nabla_c\phi \right) \right] 
 \nabla^c\phi \nabla_c\phi \right. \nonumber\\
&&\nonumber\\
& - & \left. \left[ 4+ \mbox{sign} \left( \nabla^c\phi 
\nabla_c\phi \right) \right] V(\phi) \right\} 
\,.\label{questa2} 
\end{eqnarray}
If we restrict ourselves to the case 
in which $\nabla^c\phi$ is  {\em timelike}, 
$ \nabla^c \phi  \nabla_c \phi<0$,  we have
\begin{eqnarray}
\rho & = & - \frac{1}{2} \nabla^c \phi \nabla_c \phi + 
V(\phi)  \,, \label{Delta1}\\
&&\nonumber\\
P &=&  - \frac{1}{2} \, \nabla^c \phi \nabla_c \phi - V(\phi)  
\,,\label{Delta2}
\end{eqnarray}
and
\begin{eqnarray}
&& \left( P+\rho \right) u_au_b +Pg_{ab} \\ 
\nonumber\\
&& = \nabla_a\phi  \nabla_b\phi  -\frac{1}{2}\, g_{ab} 
\nabla^c \phi \nabla_c  \phi  -Vg_{ab}  \\  
\nonumber\\
&& \equiv  T_{ab}[\phi] 
\end{eqnarray}
in  addition to $u_cu^c=-1$.  The last equation shows not 
only that a minimally coupled scalar field 
can be given a perfect fluid description, but also that any 
perfect fluid with barotropic equation of state $P=P(\rho)$ can 
be mimicked by a scalar field with appropriate potential 
$V(\phi)$. Roughly speaking, prescribing the equation of state 
$P=P(\rho)$ corresponds to assigning a suitable potential, but 
the  correspondence between equation of state and scalar 
field potential is not one-to-one \cite{Faraoni}.

The Klein-Gordon Lagrangian density for the scalar field is 
the well known \cite{Wald}
\be \label{KGlagrangian}
{\cal L}_{KG}= -\frac{1}{2}\, \nabla^c \phi \nabla_c \phi 
-V(\phi) \,,
\ee
which coincides with ${\cal L}_1 =P$ and the variation of 
the action $S_{KG}=\int d^4x \sqrt{-g} \, P$ 
with respect to $\phi$ reproduces the Klein-Gordon 
equation~(\ref{KleinGordon}), as is also well known 
\cite{Wald}. 
Let us try to adopt 
instead the other candidate Lagrangian density 
\be
{\cal L}_2=-\rho= \frac{1}{2}\, \nabla^c \phi \nabla_c \phi 
-V(\phi) \,;
\ee
then, the variational principle 
\be
\delta S_2  \equiv  \delta \int d^4 x \, \sqrt{-g} \, {\cal 
L}_2 =0
\ee
yields 
\be
 \int d^4x \, \sqrt{-g} \left[ - \left( \nabla^c 
\delta\phi \right)  \nabla_c\phi  +  \frac{ dV}{d \phi} \, \delta 
\phi \right] =0 \, .
\ee
Using the identity $
\left( \nabla^c  \delta\phi \right)  \nabla_c\phi = \nabla^c 
\left( \delta \phi \nabla_c \phi \right) -\delta \phi \Box \phi $ 
and discarding the boundary term, one obtains
\be\label{wrongKG}
\Box \phi +\frac{dV}{d\phi}=0 \,.
\ee
Eq.~(\ref{wrongKG}) is not the Klein-Gordon equation because of 
the incorrect sign of the potential derivative term. The 
difference between eq.~(\ref{wrongKG}) and the Klein-Gordon 
equation~(\ref{KleinGordon}) disappears, of course,  if 
$V=$~constant. In this 
case the potential term $-g_{ab}V(\phi)$ in the stress-energy 
tensor $T_{ab}[\phi]$ can be attributed entirely to a 
cosmological constant, {\em i.e.}, to gravity instead of matter. 
If this constant vanishes, $V\equiv 0$, then 
eqs.~(\ref{Delta1}) and (\ref{Delta2}) yield $\rho=P$ and 
eq.~(\ref{wrongKG}) coincides with the Klein-Gordon equation. 
However, there is still something wrong: the scalar field 
sourcing gravity and appearing in the total action 
\be
S_{total}=S_{gravity}+S_{matter}=\int d^4x\, \sqrt{-g}\, 
\frac{R}{2\kappa}+S[\phi]
\ee
(where $\kappa \equiv 8\pi G$)  will give the incorrect 
field equations 
\be 
R_{ab}-\frac{1}{2}\, g_{ab}R=-\kappa \, T_{ab}[\phi] 
\ee
 instead of the Einstein equations 
\be
R_{ab}-\frac{1}{2}\,  g_{ab}R=\kappa \, T_{ab}[\phi] \,.
\ee
When $V \equiv 0$, the correct 
Lagrangian density would be ${\cal L}=P=\rho$ instead of ${\cal 
L}_2=-\rho$. (Then ${\cal L}_3 \equiv -{\cal L}_2= \rho$ 
can only describe a perfect fluid with stiff equation of 
state $P=\rho$.)

The conclusion is that, for a perfect fluid, ${\cal L}_1=P$ is 
the Lagrangian density reproducing the correct equations of 
motion, 
while ${\cal L}_2=-\rho $ (or ${\cal L}_3=\rho $) provides 
incorrect equations of motion.

For completeness, we conclude this section by going back to 
eqs.~(\ref{questa1}) and~(\ref{questa2}) and considering 
the case 
of a {\em spacelike} $\nabla^c\phi$, although this choice 
obviously does not correspond to any physical fluid. If 
$ \nabla^c \phi \nabla_c\phi >0$ it is $u_a=\nabla_a \phi 
/\sqrt{ \nabla^c \phi \nabla_c\phi}$ and, using 
eqs.~(\ref{4velocity})-(\ref{Delta2}), 
\begin{eqnarray}
u^c u_c &=& 1 \, , \\
&&\nonumber\\
\rho & = &  \frac{1}{2} \nabla^c \phi \nabla_c \phi - V(\phi)  \,,\\
&&\nonumber\\
P &=&  - \frac{1}{6} \, \nabla^c \phi \nabla_c \phi - 
\frac{5}{3} \, V(\phi)  
\,.
\end{eqnarray}
Taking again ${\cal L}_1=P$ yields the incorrect field 
equation
\be
\Box\phi - 5\, \frac{dV}{d\phi}=0 \,,
\ee
while taking ${\cal L}=\pm {\cal L}_2= \mp \rho$ yields again
\be
\Box\phi + \, \frac{dV}{d\phi}=0 \,.
\ee
Hence, the Lagrangians $ {\cal L}_1, \pm {\cal L}_2$ 
all give incorrect field equations for a scalar 
field with spacelike gradient. 

Finally, we consider a {\em null} $\nabla^c \phi 
$. In this case, the purpose of  considering a scalar field 
representation of  a perfect fluid would be the modelling of the 
only perfect fluid with null four-velocity that makes sense 
physically, {\em i.e.}, a null dust with associated   
stress-energy tensor $T_{ab}=u_a u_b $ with 
 $u_cu^c=0$, describing a distribution of coherent massless 
$\phi$-waves \cite{Stephanietal, nulldust}. In  this case, 
it must be $V=0$ and  $ {\cal L}_1$ and ${\cal L}_2$, which 
are both 
proportional to $\nabla^c\phi\nabla_c \phi$, vanish 
identically and 
the usual Lagrangian density~(\ref{KGlagrangian}) 
cannot describe the null dust. In fact, a minimally coupled 
scalar field cannot work as a model of conformally invariant 
fluid such as a null dust. To model such a fluid, the 
physics of  the scalar field would need to be conformally 
invariant, which 
can only be achieved by coupling conformally the scalar to the 
Ricci curvature of spacetime $R$ via the introduction of 
the term 
$- R \phi^2 /12 $ in the action $S_{KG}$ 
\cite{ChernikovTagirov, CCJ,  Friedlander}. Moreover, the scalar 
$\phi$ must be either free ($V \equiv 0$), or have a  
quartic self-coupling $V=\lambda  \phi^4 $ 
\cite{CCJ}. In general, mimicking  a null fluid or 
an imperfect fluid with a scalar field requires that the latter  
be coupled  non-minimally  to the curvature  \cite{Madsen, 
think}.

\section{The Noether approach}

In flat spacetime there is an  independent line of approach 
to the 
Lagrangian description of a perfect fluid. The Noether theorem 
applied to the translational Killing fields of the  Poincar\'e 
group for a field $\phi$ described by the Lagrangian density 
${\cal L}[ \phi, \partial_a \phi, \eta_{ab} ]$ (where $\eta_{ab}$ 
is the Minkowski metric) yields the (independent) canonical  
energy-momentum tensor \cite{Wald}
\be
S^{ab}= \frac{ \partial {\cal L}}{\partial \left( \partial_a \phi 
\right)} \, \partial^b \phi -\eta^{ab} {\cal L} \,.
\ee
This tensor is conserved and coincides with the canonical 
$T_{ab}[\phi] $ of eq.~(\ref{tmunu}) up to a constant 
\cite{Wald}. By adopting 
\be
{\cal  L}_1 \equiv -\frac{1}{2} \eta^{ab} \partial_a\phi 
\partial_b 
\phi -V(\phi) 
\ee
one obtains 
\be
S^{ab}_{(1)}=-  \partial^a \phi \, \partial^b \phi  
+\frac{1}{2}\,  \eta^{ab} \, \partial^c \phi\partial_c\phi +V 
\eta^{ab}=-T^{ab}[\phi] \;.
\ee
As is well known, the conservation equation $\nabla^b 
T_{ab}[\phi]=0$ 
reproduces the Klein-Gordon equation. By contrast, using 
\be
{\cal L}_2 \equiv  -\frac{1}{2} \eta^{ab} \partial_a\phi 
\partial_b  \phi     + V(\phi) 
\ee
one obtains the incorrect energy-momentum tensor 
\be
S^{ab}_{(2)}=-  \partial^a \phi \, \partial^b \phi  
+\frac{1}{2}\, 
\eta^{ab}\, \partial^c \phi\partial_c\phi - V \eta^{ab}   \;,
\ee
which does not reproduce $T^{ab}[\phi]$ and the 
Klein-Gordon equation unless $V\equiv 0$.

To reiterate the argument, one can consider another situation in 
which the Noether symmetry approach is applicable: that of a 
spatially homogeneous and isotropic  
Friedmann-Lema\^{i}tre-Robertson-Walker universe with line 
element
\be
ds^2=-dt^2+a^2(t) \left( dx^2+dy^2+dz^2 \right)
\ee
in comoving coordinates (for simplicity, we consider here only 
the spatially flat metric).  The minimally coupled Klein-Gordon 
field in this spacetime depends only on the comoving time, 
$\phi=\phi(t)$, 
and its energy density and pressure are
\begin{eqnarray}
\rho(t) &=& \frac{1}{2}\, \dot{\phi}^2 + V(\phi) 
\,,\label{cosmodensity}\\
&&\nonumber\\
P(t) &=& \frac{1}{2}\, \dot{\phi}^2 - V(\phi) 
\,.\label{cosmopressure}
\end{eqnarray}
By adopting the Lagrangian density
\be
{\cal L}_1\left( a, \phi, \dot{\phi} \right) =P \,,
\ee
a suitable point-like Lagrangian is
\be
L_1={\cal L}_1 \sqrt{-g}={\cal L}_1 a^3=a^3\left( 
\frac{\dot{\phi}^2}{2}-V \right)=a^3P \,.
\ee
The Euler-Lagrange equation 
\be
\frac{d}{dt} \left( \frac{\partial L_1}{\partial 
\dot{\phi} } \right)-\frac{\partial L_1}{\partial \phi}=0
\ee
then yields the correct Klein-Gordon equation
\be  \label{correctcosmology}
\ddot{\phi}+3H\dot{\phi}+\frac{dV}{d\phi}=0 \,.
\ee
By contrast, the second point-like Lagrangian 
\be
L_2={\cal L}_2 \sqrt{-g} \, = a^3{\cal L}_2=-a^3\left( 
\frac{\dot{\phi}^2}{2}+ V 
\right)=-a^3 \rho 
\ee
yields the incorrect equation of motion
\be \label{incorrectcosmology}
\ddot{\phi}+3H\dot{\phi}- \frac{dV}{d\phi}=0 \,.
\ee

\section{Conclusions}

The duality between a minimally coupled scalar field and a 
perfect fluid is widely acknowledged, but it is not a 
complete one. Lagrangian densities which are equivalent for 
a perfect  fluid cease to be equivalent for an effective 
fluid  constructed out of a scalar field. Specifically, the 
change ${\cal L}_1=P$ to ${\cal L}_2=-\rho$, which does not 
have consequences for the equations of motion of a real 
fluid, does change the Klein-Gordon equation of motion of a 
scalar. Could this change point to the possibility that a  
scalar field exist in nature which satisfies 
eq.~(\ref{wrongKG})  instead of eq.~(\ref{KleinGordon})? In 
the cosmological literature, such a  field (satisfying 
eq.~(\ref{incorrectcosmology}) instead 
of~(\ref{correctcosmology})) is known as a {\em phantom 
field} \cite{phantom, Caldwelletal, Carrolletal}. This 
hypothetical scalar field would cause superacceleration of 
the universe ({\em i.e.}, Hubble parameter $H \equiv 
\dot{a}/a$ increasing with time, $\dot{H}>0$) and leads to 
a Big Rip singularity at a finite time in the future 
\cite{Caldwelletal, Carrolletal}. 

Phantom fields have been the subject of much attention in 
cosmology, due to repeated reports from astronomers that 
the effective equation of state parameter $w \equiv P/\rho$ 
of the cosmic fluid lies in a range which does not 
exclude, 
or even favours, $w<-1$: this is a signature of a phantom 
scalar field which causes superacceleration. There is 
consensus that phantom scalar fields are unstable, 
classically and, even more, quantum mechanically 
\cite{stability, Caldwelletal}, therefore a phantom field 
is extremely unlikely. However, a phantom 
can appear in low-energy effective actions which are 
eventually modified by higher order corrections. In the 
cosmological literature, a duality between a canonical 
scalar field and a phantom one is is obtained by changing 
the sign of the kinetic energy density term in 
eqs.~(\ref{cosmodensity}) and (\ref{cosmopressure}). Our 
discussion puts a new twist on this duality, in that a 
phantom scalar can be obtained from a canonical one by the 
exchange ${\cal L}_1 \longleftrightarrow {\cal L}_2$. It is 
presently unknown whether this exchange has a 
more fundamental meaning.

We have mentioned that a null fluid could be constructed 
using a scalar field with 
potential $V(\phi)=\lambda \phi^4 $ conformally coupled to 
the curvature and with lighlike gradient. In general, a 
non-minimally coupled 
scalar field corresponds to an imperfect fluid whose 
stress-energy tensor contains terms which can be 
interpreted as heat currents and anisotropic stresses 
\cite{Madsen}. Since a 
Lagrangian description of imperfect fluids has not yet 
been developed, we cannot comment on this aspect of the  
non-minimally coupled scalar field/imperfect fluid duality. 
It is plausible, however, 
that the obstruction to a perfect duality found for perfect 
fluids will carry over to (effective) imperfect 
fluids constructed with non-minimally coupled scalar 
fields, if they are found to admit a Lagrangian 
description.

\begin{acknowledgments}

The author thanks a referee for constructive remarks and 
the  Natural Sciences and Engineering Research Council of 
Canada for financial support. 
\end{acknowledgments}


\end{document}